# Assessing the Predictability of Social and Economic Time-Series Data: the example of crime in the UK

**Paul Ormerod** (**pormerod@volterra.co.uk**) *
and
**Laurence Smith** (**lsmith@volterra.co.uk**)

**Volterra Consulting Ltd**

The Old Power Station
121  Mortlake High Street
London SW14 8SN

* corresponding author

Paper prepared as part of the Non-Linear Modelling of Crime research project.  We are grateful to the Home Office for both financing this research, and for giving permission to allow the results to be placed in the public domain.

*Summary*: Policy targets are being set increasingly for social and economic variables in the UK. This approach requires that reasonably successful ex ante forecasts can be made. We propose a general methodology for assessing the extent to which this can be done. It has already been applied to GDP growth rate data. In this paper, we illustrate the technique with reference to property and violent crime in the UK over the post-war period. The forecasting record of the Home Office is poor. We show that this arises from an inherent property of the data, namely a lack of genuine information content. It is not possible in the current state of scientific knowledge to improve significantly on the Home Office record. We relate these findings to the econometric methodology of co-integration, and show that the correlation matrix of the data used in the Home Office models is dominated by noise.

# 1    Introduction

The concept of targeting has become increasingly important in social and economic policy in the UK, particularly since the 1997 General Election.  Targets are now set for a wide range of social and economic indicators.  This approach, to be useful, requires that accurate predictions of the variables which are targeted can be made.  We must be able to make a good assessment of the value which a targeted variable is likely to take in the absence of policy action in order to, first, set a reasonable target to be achieved and, second, to judge the success or otherwise of the policy *ex post*.

The level of crime is a particularly sensitive political issue.  But the forecasting record of the relevant government department has not been good. The UK Home Office has published details of the statistical models used as the basic input into their projections of property crime (Dhiri *et.al.* 1999).  The models use data over the post-war period from the early 1950s to the late 1990s, and the authors themselves note that the models have not performed well in the late 1990s.

The shortcomings of the models can be seen quite readily.  For example, if the burglary equation is estimated using annual data 1952-96, and 1997 is predicted, even given actual values of the explanatory variables, a fall of 4.3 per cent is projected.  The actual change was a fall of 13.6 per cent.  If we estimate the equation 1952-97, and predict 1998 given actual values of the explanatory variables, a fall of 1.1 per cent is predicted.  The actual was a fall of 5.0 per cent.  Estimating the model for theft using data 1952-96, and predicting 1997 given actual values of the explanatory variables, a rise of 2.7 per cent is predicted.  The actual was a fall of 9.5 per cent.  If we estimate the equation 1952-97, and predict 1998 given actual values of the explanatory variables, a rise of 3.5 per cent is predicted.  The actual was a fall of 1.4 per cent.

In this paper we offer an explanation for the poor forecasting record on crime.  Essentially, crime data contain insufficient information for predictions to be made which have any reasonable degree of systematic accuracy over time. It is not possible in the current state of scientific knowledge to improve the accuracy of such forecasts regardless of the particular technique or model which is used.

The methodology which we use to draw this conclusion can be applied more generally to social and economic time series to examine the likely degree of forecasting accuracy of such series over time.  Ormerod and Mounfield (2000) examine time series data on GDP in 17 Western economies, and are able to account for the persistent failures of prediction of the overall rate of growth of the economy.  The analytical technique of random matrix theory, which has extensive applications in physics, is used.

In this paper, we apply this technique to investigate the inherent degree of predictability of two important series on crime in the UK, namely domestic burglary and violent crime. Section 2 sets out the analytical framework, and section 3 shows the key results in terms of the predictability of annual changes in crime rates.  Section 4 examines the issue of building models to account for property crime over the past, and relates this to the co-integration approach of econometric theory.  Section 5 provides the conclusion.

## 2    Analysis of domestic burglary and violent crime data

Recent work by physicists (for example, Mantegna and Stanley (1999), Laloux *et.al.* (1999) and Plerou *et.al.* (1999) draws on the theoretical properties of random matrices to examine the behaviour of asset price returns in financial markets. A random matrix is a matrix whose elements are composed of independent, identically distributed random variables. The work notes standard results on the density of the eigenvalues of the correlation matrix of a random matrix (Mehta 1991) and compares these with the eigenvalues of empirical correlation matrices of asset price returns.

A reliable empirical determination of the correlation matrix is problematic. For a set of N time series of length T the correlation matrix will contain $N(N-1)/2$ distinct values (which are determined from the N time series of length T). However if T is of the same order of magnitude as N then it is reasonable to expect that the determination of the covariances will be noisy. That is to say the empirical correlation matrix is to a large extent random and its structure is dominated by noise.

For a scaled random matrix **X** of dimension T x N, (i.e where all the elements of the matrix are drawn at random and then the matrix is scaled so that each column has mean zero and variance one), then the distribution of the eigenvalues of the correlation matrix of **X** is known in the limit T, N $\rightarrow \infty$ with Q = T/N >= 1 fixed (Mehta 1991). The density of the eigenvalues, $\lambda$, is given by:

$$\rho(\lambda) = \frac{Q}{2\pi} \frac{\sqrt{(\lambda_{max} - \lambda)(\lambda - \lambda_{min})}}{\lambda} \qquad \text{for } \lambda \in [\lambda_{min}, \lambda_{max}] \qquad (1)$$

and zero otherwise, where $\lambda_{max} = \sigma^2 (1 + 1/\sqrt{Q})^2$ and $\lambda_{min} = \sigma^2 (1 - 1/\sqrt{Q})^2$ (in this case $\sigma^2 = 1$ by construction).

The eigenvalue distribution of the correlation matrices of matrices of actual data can be compared to this distribution and thus, in theory, if the distribution of eigenvalues of an actual matrix differs from the above distribution, then that matrix will not have random elements, i.e. there is structure present in the correlation matrix.

In financial markets, data is available over time for a large number of individual series. In contrast, when examining forecasts of economic growth, for example, for each country, such as the US, say, there is only one series available over time for the rate of growth of GDP. Similarly, there is only one historical realisation of, say, violent crime in the UK. In such circumstances, when just a single historical series exists, a standard method in the time series analysis of dynamic systems is to form a delay matrix from the original series (Mullin, 1993).

Let **x**(t) be a T x 1 vector of observations of the rate of growth of GDP at time t, where t runs from 1 to T. We form a delay matrix, **Z**, such that the first column of **Z** is **x**(t), the second **x**(t-1), and so on through to **x**(t-m) in the (m+1)th column. By suitable choice of m, the delay matrix can span what is usually thought of as the time period of the cycle of the economic or social data being examined.

Applying this technique to macro-economic data for the growth of output in 17 Western countries (Ormerod and Mounfield, *op.cit.*), it is found that the data contain only a small amount of genuine information, which accounts for the substantial errors observed in the track record of forecasting such series (for example, Mellis and Whittaker (1998) and Oller and Barot (1999) ).

In this particular application, the theoretical range of the eigenvalues given by (1) needs to be modified for two reasons. First, the matrices we are dealing with can be labelled "small". There are only 48 annual observations available in the data for domestic burglary and violent crime. The effect of smaller matrices is to increase slightly the maximum and decrease slightly the minimum values of the eigenvalues compared to their values in the theoretical limit. The exact distribution of eigenvalues with smaller matrices has not so far been calculated. Instead we can approximate the distribution by Monte Carlo simulation. This involves generating large numbers of random matrices of the required size and calculating the eigenvalues of their correlation matrices. We can then compare eigenvalues created from matrices generated by other sources with the eigenvalues generated from random matrices.

The second reason arises from the formation of the delay matrix, **Z.** The columns of this matrix are not strictly independent of each other. We therefore take a random series y(t) of length T and form the delay matrix **Y**, in the same way that **Z** was formed. We then calculate the eigenvalues of the correlation matrix of Y. Repeating this for a large number of random y(t), we can approximate the distribution of each of the eigenvalues. We want to establish whether it is likely that the eigenvalues derived from x(t) could have arisen from a random time series. We have the approximate distribution of eigenvalues from the y(t) and use this distribution to create a 95 percent confidence interval for each eigenvalue. If any of the eigenvalues from x(t) are outside this range we can reject the hypothesis at a 5% level that x(t) is a random time series.

The following table shows a comparison of the values of the highest and lowest eigenvalues for the theoretical limiting distribution, the simulated random matrices using data series of the same length as those of the domestic burglary and violent crime series and the simulated delay matrices, using a lag of ten years. As can be seen, the effect is to widen slightly the range between $\lambda_{max}$ and $\lambda_{min}$ compared to that of the limiting distribution.

|  | limiting distribution | simulated random matrices | 95% range of simulated random delay matrices |
| --- | --- | --- | --- |
| Highest eigenvalue | 2.29 | 2.77 | 2.65 |
| Lowest eigenvalue | 0.24 | 0.13 | 0.15 |

## 3. The results

The results are best interpreted graphically. Eigenvalues are generated from the data series being investigated[1]. These are compared to the values of the eigenvalues generated from 5000 random delay matrices. The graphs plot, as " + ", the 97.5% and 2.5% quantiles of each of the eigenvalues from the random data and plot the eigenvalues from the actual data as " o ".

Random matrix analysis was carried out on the annual percentage changes in violent crime from 1947 to 1997. It was carried out for lags of 4 to 12 years to allow for the possibility of different cycle lengths. The results showing the most structure are for lags of 5 years and are shown in the graph below. (Results for other years are available from the authors)

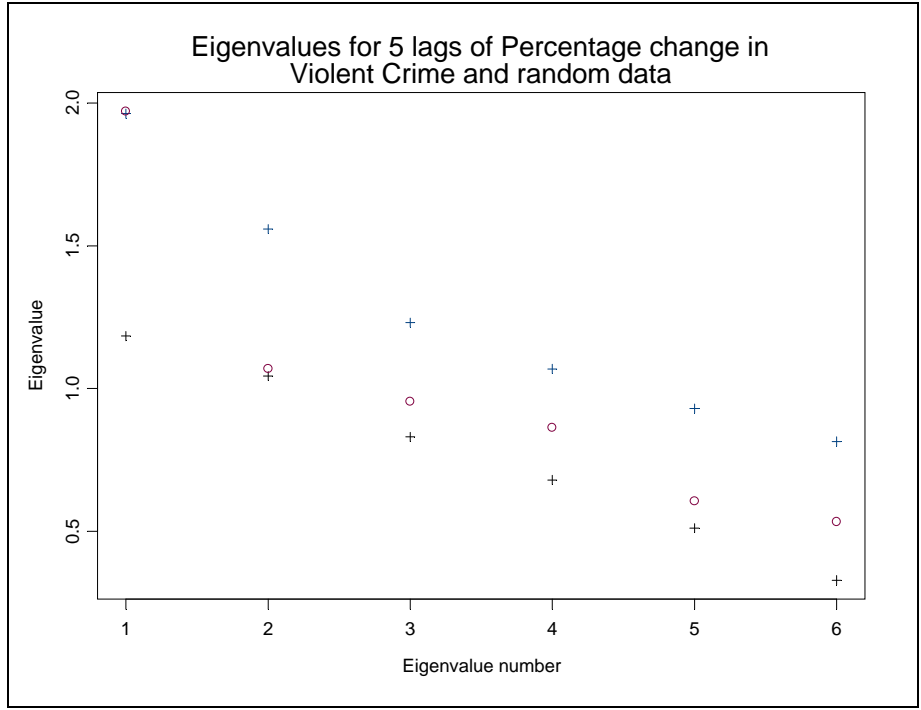

The first eigenvalue is a small amount outside the 95 percent confidence interval[2]. This means that there is a small amount of evidence for the existence of structure in the data, but there is not a great deal.

The results for domestic burglary are very similar. In this case, the results showing the most amount of structure were those for lags of 6 years, shown in the graph below, but again the deviation from the eigenvalues of the correlation matrices of random delay matrices is minimal. Again, results for different lags are available from the authors.

---

[1] the data are scaled so that each variable is normally distributed with zero mean and standard deviation of one
[2] (The "o" is above the top "+")

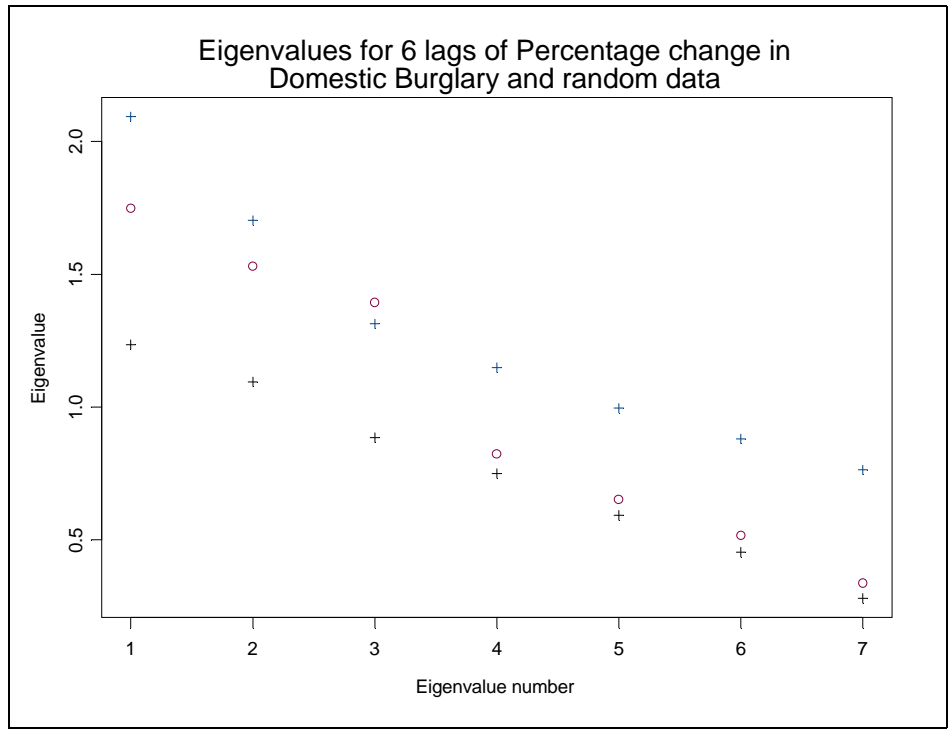

## 4. Explaining movements in crime in the past

The results above are obtained using the annual percentage changes in violent crime and domestic burglary data. Econometricians, particularly in the UK, place great emphasis on so-called co-integrating relationships (Banerjee *et.al.* 1993). These purport to establish genuine long-run relationships between the level of a variable, y(t), and the level(s) of variables, $x_i$(t), which are postulated to cause movements in y(t). The Home Office study (Dhiri *et.al., op.cit.*) reports such a relationship between burglary and two variables, the stock of consumer durable goods and the number of males aged 15 and 20 (all variables in natural logs).

It might be argued that although the annual percentage changes in a variable might be dominated by noise, over a period of years the trend in its movements could nevertheless be predicted by a co-integrating relationship. So, for example, if we could predict with reasonable accuracy the trend movements in the stock of durables and the number of young men, we could obtain satisfactory predictions of the trend in burglary.

Random matrix theory can illuminate not only this particular question, but also the extent to which it is possible to obtain useful co-integrating relationships between any given set of variables. Regression analysis relies upon the correlation matrix, and random matrix theory informs us about the relative amounts of noise and genuine information in that matrix.

Data is available (Dhiri *et.al.*) on a annual basis from 1951-98 for burglary, the stock of consumer durable goods and the number of young males. The eigenvalues of the correlation matrix of the data matrix formed with these three variables are: 2.204, 0.756 and 0.039.

The theoretical range of the eigenvalues of the correlation matrix of a matrix of the same order is given by (1) as a maximum of 1.563 and a minimum of 0.563. Simulation of 5,000 random matrices, to allow for possible small sample bias, gives a range of 1.708 to 0.399. The largest eigenvalue calculated from the actual data is clearly outside this range. The simplest hypothesis that the correlation matrix is 'pure' noise is therefore inconsistent with the calculated eigenvalue distribution

This suggests that there is some genuine information in the correlation matrix of the three variables over the 1951-98 period. However, the largest eigenvalue of this matrix is not decisively larger than the range obtained from random matrices, pointing to the presence of noise in this correlation matrix. This can be seen by examining the properties of the normalised eigenvector associated with the largest eigenvalue. Its components are: 0.649 (burglary), 0.411 (young males) and 0.640 (stock of consumer durables).

Clearly the first and last components of this eigenvector contribute most to the properties of the eigenstate (the magnitude of each component of the eigenvector corresponds to the contribution of that component – in this instance burglary, stock of consumer durables and number of young males – to the eigenvector).

This may be quantified via the *Inverse Participation Ratio*, which is defined as (Mantegna and Stanley, *op.cit.*)

$$I = \sum_{i=1}^{N} (u_i)^4$$

where $u_i$ is the i component of the eigenvector $\vec{u}$ and N is the number of components of the eigenvector (in this case N = 3). I therefore measures the relative contribution of each eigenvector component to the overall eigenstate. The reciprocal of I gives the number of eigenvector components significantly different from zero. If all N components were contributing equally then we would expect $1/I = N$. If there were only one component contributing to the eigenvector then we would expect $1/I = 1$ (this is readily demonstrated by constructing a dummy eigenvector and calculating I by hand).

For this particular eigenvector we have I = 0.374 and so there are 2.67 eigenvector components contributing to this eigenvector. This result therefore indicates that not all of the 3 components are contributing equally to the eigenvector, and that two of the eigenvector components have a greater correlation with one another than with the other eigenvector component. Visual inspection of the eigenvector demonstrates that burglary is more highly correlated with the stock of consumer durables that it is with the number of young males.

Information derived from random matrix theory therefore suggests two things. First, there is genuine information in the empirical correlation matrix, but there is also a distinct degree of noise. This implies that this correlation matrix will not be time invariant. Second, that burglary is related to the stock of durables more reliably than it is to the number of young men.

The first of these points has implications for both the ability to use co-integrating relationships for predicting trends, and for the ability to estimate such relationships over the past which are time-invariant. Co-integrated relationships are estimated using classical least squares regression[3], and Table 2 does this for the relationship reported by the Home Office for the burglary data. The estimation period begins in 1951, but instead of just estimating it using the whole of the data through to 1998, we first of all use data from 1951-80 then 1951-81, 1951-82, and so on through to 1951-98. The model reported by Dhiri *et.al.* satisfies a battery of conventional statistical tests, including those for parameter stability. Table 2 shows quite clearly, however, that the parameters are not stable in any useful sense of the word.

---

[3] or maximum likelihood regression

| Table 2: Estimated coefficients on co-integrated relationship reported for burglary by the Home Office ||| 
| | Coefficient on: ||
| **Estimation period** | **Stock** | **Males** |
| 1951-80 | 1.508 | 1.747 |
| 1951-81 | 1.509 | 1.747 |
| 1951-82 | 1.521 | 1.756 |
| 1951-83 | 1.535 | 1.754 |
| 1951-84 | 1.554 | 1.754 |
| 1951-85 | 1.573 | 1.738 |
| 1951-86 | 1.607 | 1.697 |
| 1951-87 | 1.614 | 1.688 |
| 1951-88 | 1.524 | 1.823 |
| 1951-89 | 1.464 | 1.921 |
| 1951-90 | 1.550 | 1.779 |
| 1951-91 | 1.682 | 1.558 |
| 1951-92 | 1.826 | 1.299 |
| 1951-93 | 1.927 | 1.113 |
| 1951-94 | 1.961 | 1.051 |
| 1951-95 | 1.973 | 1.029 |
| 1951-96 | 1.963 | 1.048 |
| 1951-97 | 1.923 | 1.128 |
| 1951-98 | 1.875 | 1.221 |

This illustrates the points made above from a random matrix theory analysis of the data very clearly. First, and most importantly, the long-run relationship between burglary and the stock of durables and males is not in any meaningful sense time-invariant. The coefficient of males in particular is very sensitive to the estimation period use, with the addition of just a single observation sometimes leading to large shifts in its estimated value. The coefficient on the stock of durable goods is more stable, but even this varies between 1.508 and 1.973. The second point made above is precisely that we would expect greater stability of the stock of goods coefficient than of the males coefficient.

In other words, it is not really possible to use this particular co-integrating relationship to carry out reasonably accurate trend forecasts over a period of years. Such forecasts are likely to be somewhat more accurate over time than are predictions of year-on-year changes, but cannot claim a high degree of potential reliability.

## 5. Conclusion

In summary, the forecasting record over time of the annual rate of changes in domestic burglary and violent crime in the UK will be poor, no matter what economic theory or statistical technique is used to generate them. This is due to inherent characteristics of the data. For the periods over which any regularity of behaviour of the crime data might be postulated to exist, the genuine information content of correlations over time in the data is low.

Taking as an example the Home Office econometric model for burglary, we find that there is more genuine information in the correlation matrix of the level of crime - rather than its annual rate of change - and the factors which are hypothesised to cause it. Long run relationships which are identified between levels of variables may be of some use in identifying the trend in crime, but even so the correlation matrix contains a substantial degree of noise, and such relationships will not be time-invariant.

The methodology we propose here can be used more generally on social and economic time series to data to assess both the stability of the empirical correlation matrix of the relevant data, and the inherent degree of predictability of the data.

# References

A.Banerjee, J.Dolado, J.Galbraith and D.Hendry, *Co-integration, Error-Correction, and the Econometric Analysis of Non-Stationary Data*, OUP, Oxford, (1993)

S.Dhiri, S.Brand, R.Harries and R.Price, 'Modelling and Predicting Property Crime Trends', *Home Office Research Study* **198**, Home Office, London (1999)

L. Laloux, P. Cizeau, J-.P. Bouchaud and M. Potters, 'Noise Dressing of Financial Correlation Matrices', Phys. Rev. Lett **83**, 1467 (1999)

R. N. Mantegna and H. E. Stanley, *An Introduction to Econophysics,* Cambridge University Press (1999) and references therein

M. L. Mehta, *Random Matrices*, Academic Press, Boston (1991)

C.Mellis and R.Whittaker, 'The Treasury forecasting record: some new results', *National Institute Economic Review*, no. 164, pp.65-79, . (1998)

T.Mullin, 'A dynamical systems approach to time series analysis', in T.Mullin, ed., *The Nature of Chaos*, Oxford Scientific Publications, (1993)

P.Ormerod and C.Mounfield, 'Random Matrix Theory and the Failure of Macro-economic Forecasting', *Physica* **A** 280 (2000) 497-504

V. Plerou, P. Gopikrishnan, B. Rosenow, L. A. Nunes Amaral and H. E. Stanley, 'Universal and Non-universal Properties of Cross Correlations in Financial Time Series', *Phys. Rev. Lett* **83**, 1471 (1999)